\documentclass[twocolumn,showpacs,amsmath,amssymb,bibnotes]{revtex4}
\usepackage{bm}
\usepackage{graphicx,psfig,epsfig,subfigure,afterpage}
\usepackage{amsfonts}
\newcommand{\bw}{\begin{widetext}}
\newcommand{\ew}{\end{widetext}}
\newcommand{\gt}{\gdot\,t}
\newcommand{\Cg}{C(t,\gdot)}
\newcommand{\Rg}{\chi(t,\gdot)}
\newcommand{\be}{\begin{equation}} \newcommand{\ee}{\end{equation}}
\def\(#1){(\ref{#1})} \newcommand{\bea}{\begin{eqnarray}}
  \newcommand{\eea}{\end{eqnarray}} 
\newcommand{\deriv}[1]{{\partial\over\partial{#1}}}
\newcommand{\lav}{\left\langle} \newcommand{\rav}{\right\rangle}
 
\newcommand{\av}[1]{\left\langle #1 \right\rangle}
 \newcommand{\tw}{t_{\rm w}}
\newcommand{\dt}{t\!-\!t_{\rm w}} 
\newcommand{\twlim}{t_{\rm w}\to\infty}

\newcommand{\distrm}{\sigma(m|E)}
\newcommand{\trapmean}{\overline{m}(E)}

\newcommand{\trapvar}{\Delta^2(E)}

 \newcommand{\C}{C(t-t_{\rm w})}
\newcommand{\R}{\chi(t-t_{\rm w})} \newcommand{\CC}{C(t,t_{\rm w})}
\newcommand{\RR}{\chi(t,t_{\rm w})} \newcommand{\XX}{X(t,t_{\rm w})}
\newcommand{\teff}{T_{\rm eff}} \newcommand{\tg}{T_{\rm g}}
\newcommand{\gdot}{\dot{\gamma}} 
\newcommand{\etal}{{\it et al.}}

\newcommand{\Yss}{Y_\infty}
\newcommand{\PElg}{P_{\infty}(E,l)}
\newcommand{\PEg}{P_{\infty}(E)}
\newcommand{\versus}{{\it vs.}}

\begin{document}

%%%%%%%%%%%%%%%%%%%%%%%%%%%%%%%%%%%%%%%%%%%%%%%%%%%%%%%%%%%%%%%%%%%%%%%%%%%%%
\title{Equivalence of driven and ageing fluctuation-dissipation
  relations in the trap model} \author{S. M. Fielding}
\email{physf@irc.leeds.ac.uk} \affiliation{Polymer IRC and Department
  of Physics \& Astronomy, University of Leeds, Leeds LS2 9JT, United
  Kingdom} \author{P. Sollich} \email{p.sollich@mth.kcl.ac.uk}
\affiliation{Department of Mathematics, King's College London, Strand,
  London, WC2R 2LS, United Kingdom} \date{\today}
\begin{abstract}
  We study the non-equilibrium version of the fluctuation dissipation
  (FD) relation in the glass phase of a trap model that is driven into
  a non-equilibrium steady state by external shear.  This extends
  our recent study of ageing FD relations in the same model, where we
  found limiting, observable independent FD relations for ``neutral''
  observables that are uncorrelated with the system's average energy.
  In this work, for such neutral observables, we find the FD relation
  for a stationary weakly driven system to be the same, to within
  small corrections, as for an infinitely aged system. We analyse the
  robustness of this correspondence with respect to non-neutrality of
  the observable, and with respect to changes in the driving
  mechanism.
\end{abstract}
\maketitle

%%%%%%%%%%%%%%%%%%%%%%%%%%%%%%%%%%%%%%%%%%%%%%%%%%%%%%%%%%%%%%%%%%%%%%%%%%%%%%
%     PACS numbers: explanations and other possible choices                  %
%       02.50.Ey (Stochastic processes)                                      %
%       05.20.-y (Classical stat mech)                                       %
%       05.40.-a (fluctuations, random processes, noise and Brownian motion) %
%       05.70.Ln (non eqbm and irrev TD)                                     %
%       61.43.Fs (Glasses)                                                   %
%%%%%%%%%%%%%%%%%%%%%%%%%%%%%%%%%%%%%%%%%%%%%%%%%%%%%%%%%%%%%%%%%%%%%%%%%%%%%%

\section{Introduction}
\label{sec:intro}

Glasses relax very slowly at low temperatures. They therefore remain
out of equilibrium for long times and exhibit ageing
\cite{BouCugKurMez98}: the time-scale for response to an external
perturbation (or for the decay of correlations) increases with the
``waiting time'' $\tw$ since the system was quenched to the low
temperature, and thus eventually far exceeds the experimental
time-scale. Time translational invariance (TTI) is lost.  As a result
of this dynamical sluggishness, glassy systems are highly susceptible
to external driving, even when the driving rate $\gdot$ is small.  One
example of $\gdot$ is shear rate in a rheological system.  Typically,
steady driving interrupts ageing and restores a non-equilibrium steady
(TTI) state in which the time-scale defined by the inverse driving
rate plays a role analogous to the waiting time $\tw$ of the ageing
regime~\cite{CugKurLeDouPel97,HerGriSol86,CriSom87,Horner96,Thalmann98,BerBarKur00,SolLeqHebCat97}.

Ageing and driven glasses in general violate the equilibrium
fluctuation dissipation theorem (FDT)~\cite{Reichl80}. Consider the
autocorrelation function for a generic observable $m$, defined as
$\CC=\lav m(t)m(\tw)\rav -\lav m(t) \rav \lav m(\tw) \rav$.  The
associated step response function $\RR=\int_{\tw}^t\!  dt'\, R(t,t')$
tells us how $m$ responds to a small step $h(t)=h\Theta(\dt)$ in its
conjugate field $h$.  In {\em equilibrium}, $\CC=\C$ by TTI (similarly
for $\chi$), and the FDT reads $-\deriv{\tw} \R = R(t-\tw) =
\frac{1}{T}\deriv{\tw}\C$ where $R(t-\tw,\tw)=\left.\frac{\delta \lav
    m(t)\rav}{\delta h(\tw)}\right|_{h=0}$ is the impulse response
function and $T$ is the thermodynamic temperature. (We set $k_{\rm
  B}=1$.)  A parametric `FD plot' of $\chi$ vs.\ $C$ is thus a
straight line of slope $-1/T$. 

Out of equilibrium, violation of FDT is
measured by an FD ratio, $\XX$, defined
through~\cite{CugKur93,CugKur94}
\be
\label{eqn:non_eq_fdt}
-\deriv{\tw} \RR= R(t,\tw) = \frac{{\XX}}{T}\deriv{\tw}\CC.
\ee
In ageing systems, violation ($X\neq 1$) can persist even at long
times, indicating strongly non-equilibrium behavior even though
one-time observables such as entropy may have settled to essentially
stationary values.  Similarly, driven glassy systems can violate FDT
even in the limit of weak driving.

Remarkably, the FD ratio in several ageing mean field
models~\cite{CugKur93,CugKur94} assumes a special form at long times.
Taking $\tw\to\infty$ and $t\to\infty$ at constant $C=C(t,\tw)$,
$X(t,\tw)\to X(C)$ becomes a (non-trivial) function of the single
argument $C$. If the equal-time correlator $C(t,t)$ also approaches a
constant $C_0$ for $t\to\infty$, it follows that
\be
\chi(t,\tw)=\frac{1}{T}\int_{C(t,\tw)}^{C_0}.
\!dC\, X(C).
\label{mf_limit}
\ee
Graphically, this limiting non-equilibrium FD relation is obtained by
plotting $\chi$ \versus\ $C$ for increasingly large times. From the
slope $-X(C)/T$ of the limit plot an effective
temperature~\cite{CugKurPel97} $\teff(C)=T/X(C)$ can be defined.
Throughout this paper, we absorb the factor $T$ into the response
function so that an equilibrium FD plot has slope $-1$.

In the most general ageing scenario, a system displays dynamics on
several characteristic time scales, each with its own functional
dependence on $\tw$. If these time scales become infinitely separated
as $\twlim$, they form a set of distinct `time sectors'. In mean
field, $\teff(C)$ is {\em constant} within each
sector~\cite{CugKur94}, and {\em independent} of the observable $m$
used to construct the FD plot.  These properties have also been
observed in some lower dimensional (non mean field)
systems~\cite{Barrat98,MarParRicRui98}.

Cugliandolo \etal~\cite{CugKurPel97} proposed that an equivalent
limiting non-equilibrium FD relation should hold in slowly {\em
  driven} glassy systems ($\gdot\to 0$) and that the corresponding
effective temperatures $\teff(C,\gdot\to 0)$ and $\teff(C,\twlim)$
should coincide. Although this is believed to apply widely among
driven glasses, the evidence supporting it, to date, is limited to the
two detailed studies of Berthier, Barrat \etal, in mean
field~\cite{BerBarKur00} and in simulations of sheared Lennard Jones
particles, initially in Ref.~\cite{BarBer01} and later, with a study
of observable independence, in Refs~\cite{BerBar02,BerBar02b}.  We
note that FDT has also been studied in a driven phase separating model
($T<T_c$ where $T_c$ is the critical temperature) with $N$
non-conserved order parameters in the large $N$
limit~\cite{CorGonLipZan01}, although here the drive does not
interrupt ageing.  In this study, in the timesector with stationary
dynamics, the FD plot is of trivial equilibrium form for $\gdot=0$ but
a non-trivial curve for $\gdot>0$; in the ageing sector, the response
function is constant, giving a flat FD plot.

In this paper, therefore, we study FDT in the driven regime of
Bouchaud's trap
model~\cite{Bouchaud92,MonBou96,RinMaaBou00,SolLeqHebCat97}, for which
correlation and response functions can be calculated exactly. This
will allow a comparison with our recent study~\cite{FieSol02} of the
ageing trap model, where we found limiting FD relations that are
observable independent for ``neutral observables'' that are
uncorrelated with the system's energy.  This is consistent with the
mean field work, for which observables are usually defined in terms of
random couplings, uncorrelated with the average energy. (In coarsening
models, similar arguments have been used to exclude observables
correlated with the order parameter~\cite{Barrat98}.)  Surprisingly,
however, we found the FD plot to be a continuous curve even though the
model has just one time sector, with relaxation times $O(\tw)$.
Although this finding is apparently at odds with the mean field
predictions, it is likely to result from the fact that the trap model
has a broad distribution of relaxation times (all within its single
time sector). We return to this point in the conclusion.

In what follows, our central result will be that {\em the same}
non-trivial FD relation is found (to within logarithmic corrections),
even when the trap model is weakly driven according to the mechanism
proposed in Ref.~\cite{SolLeqHebCat97}.  Although the curvature of the
FD plot obviously excludes a constant effective temperature, our
finding is still consistent with Cugliandolo \etal's predictions, in
so far as {\em the relationship between correlation and response is
  the same for aged and weakly driven glasses}. This finding is
non-trivial since the shapes of the relaxation spectra for the ageing
and weakly driven trap models differ strongly from each other.

We start (Sec.~\ref{sec:trap}) by defining the trap model and
summarising the ageing FD predictions of Ref.~\cite{FieSol02}. We then
(Sec.~\ref{correlator}) derive exact expressions for the
autocorrelation and response functions for an arbitrary observable $m$
in the steadily driven regime. Using these, we calculate the limiting
driven FD relation, for neutral observables that are uncorrelated with
the average energy. We show that this relation is the same (to within
logarithmic corrections) as its ageing counterpart
(Sec.~\ref{sec:plots}). We then (Sec.~\ref{sec:otherdrive}) consider
robustness of this correspondence with respect to (i) changes in the
driving mechanism and (ii) non-neutrality of the observable, before
concluding.

\section{The trap model; driving}
\label{sec:trap}

\begin{figure}[h]
\centerline{\psfig{figure=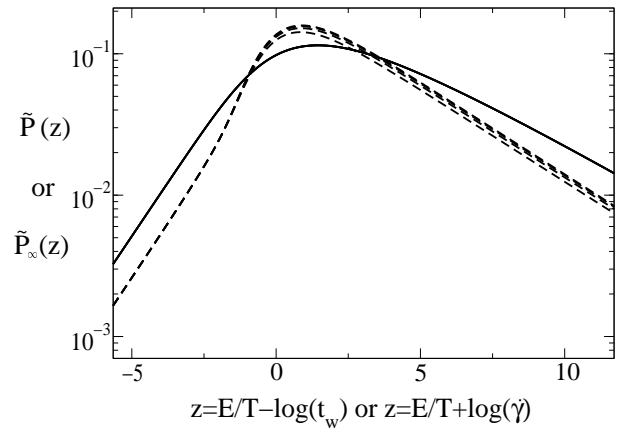,width=8cm}}
\caption{Scaled energy distributions for (i) the ageing model at  $\tw=10^{3},\,10^4,\,10^5,\,10^6$ (dashed lines) and (ii) the driven model for $\gdot=10^{-3},\,10^{-4},\,10^{-5},\,10^{-6}$ (solid lines; different $\gdot$ values indistinguishable). The temperature $T=0.3$.
\label{fig:distr} } 
\end{figure}

The trap model~\cite{Bouchaud92} comprises an ensemble of uncoupled
particles exploring a spatially unstructured landscape of (free)
energy traps by thermal activation. The tops of the traps are at a
common energy level and their depths $E$ have a `prior' distribution
$\rho(E)$ ($E>0$).  A particle in a trap of depth $E$ escapes on a
time scale $\tau(E)=\tau_0\exp({E}/{T})$ and hops into another trap,
the depth of which is drawn at random from $\rho(E)$.  The
probability, $P(E,t)$, of finding a randomly chosen particle in a trap
of depth $E$ at time $t$ thus obeys
\be
\label{eqn:trap_eom}
\partial_tP(E,t)=-\tau^{-1}(E)P(E,t)+Y(t)\rho(E)
\end{equation}
in which the first (second) term on the RHS represents hops out of
(into) traps of depth $E$, and $Y(t)$ $=$ $\av{\tau^{-1}(E)}_{P(E,t)}$ is
the average hopping rate. 

For the specific choice of prior distribution
$\rho(E)\sim\exp\left(-{E}/{\tg}\right)$, the model shows a glass
transition at a temperature $\tg$.  This is seen as follows.  At a
temperature $T$, the equilibrium state (if it exists) is $P_{\rm
  eq}(E)\propto\tau(E)\rho(E)\propto\exp\left({E}/{T}\right)
\exp\left(-{E}/{\tg}\right)$. For temperatures $T\le \tg$ this is
unnormalizable, and cannot exist; the lifetime averaged over the
prior, $\av{\tau}_{\rho}$, is infinite.  Following a quench to
$T\le\tg$, the system never reaches a steady state, but instead ages.
This can be seen from the time evolution of $P(E,\tw)$, which can be
obtained exactly from Eqn.~\ref{eqn:trap_eom} (with $Y(\tw)$
determined self consistently by enforcing normalisation of
$P(E,\tw)$~\cite{Bouchaud92,MonBou96}). At large times $\tw\to\infty$
a scaling limit is reached in which $P(\tau,\tw)=[T/\tau(E)]P(E,\tw)$
is concentrated entirely on traps of lifetime $\tau=O(\tw)$: the
scaling distribution $\tilde{P}(z)=T P(E)$ where $z=E/T-\log(\tw)$ is
shown in Fig.~\ref{fig:distr}.  The model thus has just one
characteristic time scale, which grows linearly with the age $\tw$.
(In contrast, for $T>\tg$ all relaxation processes occur on time
scales $O(\tau_0)$.) In what follows, we rescale all energies such
that $\tg=1$, and times so that $\tau_0=1$.

Driving was first incorporated into the model in order to study the
rheology of ``soft glassy materials''. Although we are not directly
interested in rheology here, we use the same driving rules which are
defined as follows. Each particle is assigned its own local elastic
``strain'' $l$, with a corresponding ``stress'' $kl$. Each time the
particle hops, $l$ is set to zero. Between hops, $\dot{l}=\gdot$ where
$\gdot$ is the rate of external driving (``straining''). A particle in
a trap of depth $E$ strained by $l$ sees a reduced effective energy
barrier $E-\tfrac{1}{2}kl^2$, so that
\be
D_t P(E,l,t) = - \tau^{-1}(E)e^{kl^2/2T}\,P + Y(t)\,\rho(E)\delta(l).
\label{eqn:driventrap}
\ee
In this equation, $\tau^{-1}(E)e^{kl^2/2T}$ is the strain-enhanced
counterpart of the ``bare'' activation rate $\tau^{-1}(E)$ defined
above for the undriven model, and $D_t$ is the convected derivative
$D_t=\partial_t+\gdot\partial_l$. Eqn.~\ref{eqn:driventrap}
(integrated on $l$) reduces to Eqn.~\ref{eqn:trap_eom} for $\gdot=0$,
as required. In the following we rescale $l$ such that $k=1$.

For steady driving ($\gdot={\rm const.}$) ageing is interrupted, and a TTI
steady state is restored. To write down the steady state distribution,
define $S(l_1, l_2, E)$ as the probability for a particle that
starts off with strain $l_1$ and in a trap of depth $E$ not to hop until its
strain has reached $l_2$. From Eqn.~\ref{eqn:driventrap}, this is
\be
S(l_1, l_2, E) = 
\exp\left[-\frac{1}{\tau(E)\gdot}\int_{l_1}^{l_2}
dl\,\exp\left(\frac{l^2}{2T}\right)\right]
\ee
In terms of this quantity, the steady state distribution of
Eqn.~\ref{eqn:driventrap} can then be written as
\be
\PElg=\frac{\Yss}{\gdot}\rho(E) S(0,l,E).
\label{eqn:steady}
\ee
Here $\Yss\sim \gdot^{1-T}$ is the steady state average hopping rate,
which can be determined self consistently by enforcing normalisation
of $\PElg$. (This is most conveniently done by changing variables to
$\tau=\exp(E/T)$ which gives $\rho(\tau)\sim\tau^{-1-T}$.) In the
limit $\gdot\to 0$, the energy distribution $\PEg=\int dl\,\PElg$
approaches a scaling limit in which all relaxation times are
$O(1/\gdot)$. The scaling distribution $\tilde{P}_\infty(z)=T P_\infty
(E)$ where $z=E/T+\log(\gdot)$ differs strongly from its ageing
counterpart $\tilde{P}(z)$, as shown in Fig.~\ref{fig:distr}.

\section{Correlation and response}
\label{correlator}

FDT can be studied by assigning to each trap, in addition to its depth
$E$, a value for an (arbitrary) observable $m$~\cite{FieSol02}.
%By analogy with spin models we refer to $m$ as magnetization.  
The trap population is then characterized by the joint prior
distribution $\distrm\rho(E)$, where $\distrm$ is the distribution of
$m$ across traps of given fixed energy $E$. The dynamics then obey
\bw
\be
D_t P(E,m,l,t) = - \tau^{-1}(E,m)e^{kl^2/2T}\,P + Y(t)\,\rho(E)\delta(l)\distrm
\label{eqn:fdtdynamics}
\ee
\ew
where the activation times are modified by a small field $h$ conjugate
to $m$ as $\tau(E,m)=\tau(E)\exp\left(mh/T\right)$. This particular
form of $\tau(E,m)$ is one of several possible choices that all
maintain detailed balance under zero-driving
conditions~\cite{RinMaaBou00,BouDea95}.  We adopt it because, in the
spirit of the unperturbed model ($h=0$), it ensures that the jump rate
between any two states depends only on the initial state, and not the
final one.

In Ref.~\cite{FieSol02} we derived exact expressions for the two-time
autocorrelation and step response functions, $\CC$ and $\RR$ in the
ageing regime ($\tw\to\infty$, $t\to\infty$ at fixed $t/\tw$) of the
undriven model ($\gdot=0$), following a quench into the glass phase at
time $\tw=0$ from an initially infinite temperature.  Each comprises
two components that depend separately upon the functional forms of the
mean, $\trapmean$, and variance, $\trapvar$, of the distribution
$\distrm$. For the purposes of this paper we are interested only in
observables with $\trapmean=0$ and (within these) mainly the
neutral observable for which the variance is uncorrelated with energy,
$\trapvar={\rm const.}\equiv 1$.  In this case, in the simultaneous
limit $\twlim$ with $t\to\infty$, $\CC$ and $\RR$ depend on time only
through the scaling variable $(t-\tw)/\tw$ as shown in
Fig.~\ref{fig:neutral}a,b. The corresponding FD plot is shown in
Fig.~\ref{fig:neutral}c.

In the steadily driven regime, TTI is restored: $C$ and $\chi$ do not
depend explicitly upon the waiting time $\tw$ but only on the
measurement interval $t-\tw$, so we set $\tw=0$ without loss of
generality.  For observables with $\trapmean=0$, the autocorrelation
function is exactly
\be
\Cg=\int_{\gt}^{\infty}dl\int_0^\infty dE\;\trapvar \PElg.
\label{eqn:correlator}
\ee
This can be understood as follows.  When any particle hops, its new
value of $m$ is uncorrelated with the old one. At time $t$, therefore,
only those particles that have not hopped since $t=0$ can contribute
to the correlator, with weight $\trapvar$. The fraction of such
particles which had strain $l$ and trap depth $E$ at time $t=0$ is
$\PElg S(l,l+\gdot t,E)$. From Eqn.~\ref{eqn:driventrap} this equals
$P_{\infty}(E,l+\gdot t)$ and integration on $l$ and $E$ gives the
result, Eqn.~\ref{eqn:correlator}. [Alternatively,
Eqn.~\ref{eqn:correlator} can be understood by recalling the driving
dynamics: upon any hop, each particle resets its local strain to zero;
between hops, the local strain affinely follows the applied one.
Therefore, the particles that have not hopped since $t=0$ are just
those that have strains $l\ge \gdot t$.]

The corresponding switch-on response function has contributions from both
hopped and unhopped particles.  The contribution from hopped particles
can be expressed as an integral over the last hop-time of each
particle, $t'$:
\bw
\be
\label{eqn:Rhop}
\chi_{\rm hop}(t,\gdot)=\partial_h|_{h=0} \int_{-\infty}^\infty dm
\int_0^\infty dE \int_0^t dt'\, Y(h,t') m
\distrm\rho(E)\exp\left[-\frac{1}{\gdot\tau(E,m)}\int_0^{\gdot(t-t')} ds
    \exp\left(\frac{s^2}{2T} \right) \right].  \ee \ew
In this expression, $Y(h,t)=\Yss+O(h)$ is the fraction of particles
that last hopped at time $t'$. Of these, a proportion $\distrm\rho(E)$
chose energy $E$ and magnetization $m$.  The subsequent survival
probability over the interval $t'=0\ldots t$ is encoded by the
exponential factor. In principle, the differentiation on $h$ has two
contributions: one from the factor $\tau(E,m)$ in the exponential, and
another from $Y(h,t)$.  However the second gives zero, since
$\int_{-\infty}^\infty m\distrm=0$ for the zero-mean variables
considered here.  Adding the contribution from particles that have
not hopped since $t=0$, and doing some manipulation, we find
finally the exact result
\bw 
\be
\Rg=\int_0^\infty\,dE\int_0^\infty\,dl\,\frac{\trapvar}{\gdot\tau(E)}
\left[\PElg-P_{\infty}(E,l+\gdot t)\right] 
\int_0^l\,ds\,\exp\left(\frac{s^2}{2T}\right)
\label{eqn:response}
\ee 
\ew
(into which we have absorbed a factor $T$, as described above).

\section{FD plots}
\label{sec:plots}

\begin{figure}[h]
  \centerline{\psfig{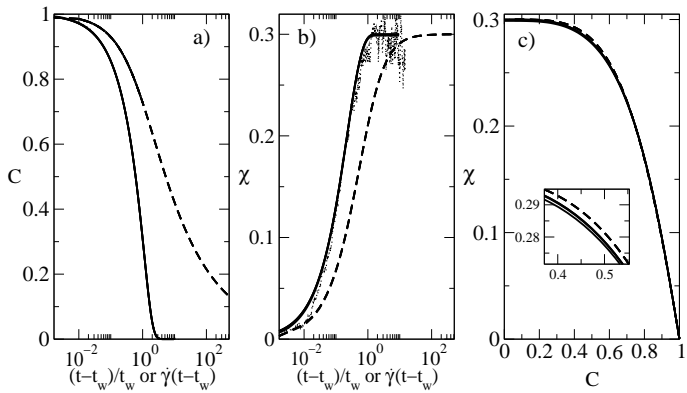}}
\caption{a) Correlator and b) response \versus\ scaled time for the
neutral observable $\trapvar=1$ in the driven model (solid lines) and
ageing model (dashed lines), calculated from the exact analytical
expressions in the text and in Ref.~\cite{FieSol02}. For the ageing
case,   waiting times $\tw=10^{3},\,10^4,\,10^5,\,10^6$  are shown
(but are indistinguishable from each other); for the driven case
shear rates 
  $\gdot=10^{-3},\,10^{-4},\,10^{-5},\,10^{-6}$ are shown (also
  indistinguishable).  As an independent check, we also show the
  driven correlator and response at $\gdot=10^{-3}$ calculated from
  waiting time Monte Carlo simulation. For $C$ this is
indistinguishable from the 
  exact results for $C$ in a); for $\chi$ it appears as the jagged line in
  b).  (c) FD plots of correlator \versus\ response for the driven
  case (solid lines) and ageing case (dashed lines) constructed from
  the exact results of (a,b); the inset is zoomed on a small region
  of the main plot. For the driven case (solid lines), driving rate
  decreases downwards at fixed $C$. The temperature $T=0.3$.
\label{fig:neutral} } 
\end{figure}

Using the exact expressions of Eqns.~\ref{eqn:correlator}
and~\ref{eqn:response}, we calculated $\Cg$ and $\Rg$ numerically for
the neutral observable, $\trapvar=1$. As an independent check, we
calculated each quantity by direct simulation, using a waiting time
Monte Carlo technique. The results are shown in
Fig.~\ref{fig:neutral}a,b. In the limit $\gdot\to 0$, $t \to 0$ at
fixed $\gdot t$, $\Cg$ and $\Rg$ depend on $\gdot$ and $t$ only
through the scaling variable $\gdot t$.  Analytically, this can be
seen for the correlator by substituting Eqn.~\ref{eqn:steady} into
Eqn.~\ref{eqn:correlator} and integrating on $dE$ by changing
variables to $\tau$, as described above. The $\gdot$ dependence from
the integral exactly cancels the prefactor $\Yss/\gdot$ so that the
only dependence on $\gdot$ and $t$ appears through the scaling
variable $\gdot t$ in the limit of the integral on $l$. A similar
argument applies to the response function.

The scaling functions $C(\gdot t)$ and $\chi(\gdot t)$ both differ
strongly from their ageing counterparts (compare the solid and dashed
lines in Fig.~\ref{fig:neutral}a,b).  This is to be expected, due to
the obvious difference between the (scaled) energy distributions of
the driven and undriven models.  Remarkably, however, the driven FD
relation $\chi(C)$ is strikingly similar to its undriven counterpart
(Fig.~\ref{fig:neutral}c).  Both start with a slope $-X(C=1)\equiv
\chi'(C=1)=-1$ (thus reproducing the equilibrium FD form in this
limit) and finish with a slope $\chi'(C=0)=0$ at an intercept
$\chi(C=0)=T$. (We have confirmed these features analytically as well
as numerically.) Even between these limits, these is little
discernible difference between the ageing and driven FD plots.  This
non-trivial result is consistent with the predictions of Cugliandolo
\etal, that the relationship between correlation and response should
be the same in weakly driven and old ageing glassy systems.

In the inset of Fig.~\ref{fig:neutral}c, we show an expanded region of
the main FD plot.  Despite the striking similarity of the ageing and
driven FD relations, our numerics do nonetheless suggest a small
discrepancy. To investigate this further, we examined the behaviour of
$X\equiv-\chi'(C)$ in the limit $C\to 0$. 
%It is in this regime that we
%expect any differences between the ageing and driven plots to be most
%pronounced, because the differences between the ageing and driven
%relaxation spectra are most manifest in the large-$E$ tails of the
%distributions shown in Fig.~\ref{fig:distr}.  
Setting $(t-\tw)/\tw=u$ for the ageing model, we found $C\sim u^{-T}$
and $X\sim u^{-1}$, hence $X\sim C^{1/T}$ as $u\to\infty$. Setting
$\gdot t=v$ for the driven model, we found $C\sim v^{T-1}\exp(-v^2/2)$
and $X\sim \exp[-v^2/2T]$, hence $C\sim X^T [\log(1/X)]^{(T-1)/2}$ as
$v\to\infty$. Therefore, the driven and ageing FD plots are indeed
equivalent in the limit $C\to 0$, but only to within minor logarithmic
corrections. This explains the slight discrepancy seen in our
numerical data.

\section{Robustness under change of driving mechanism; non-neutral observables}
\label{sec:otherdrive}

\begin{figure}[h]
\centerline{\psfig{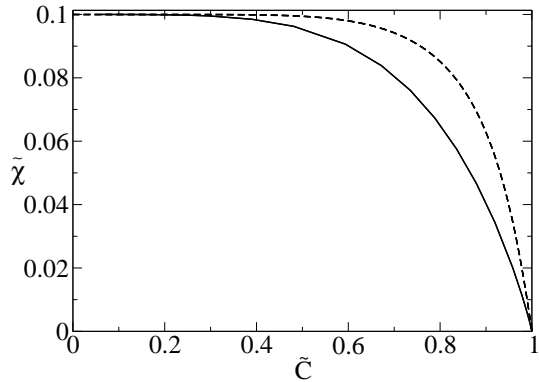}}
\caption{FD plots for (i) the ageing model with  $t=10^{6},\,10^7$ (dashed lines; different $t$ values indistinguishable) and (ii) the driven model with $\gdot=10^{-3},\,10^{-4},\,10^{-5},\,10^{-6}$ (solid lines; different $\gdot$ values indistinguishable) for a non-neutral observable with  $\trapmean=0$, $\trapvar=\exp(nE/T)$ for $n=0.2$. The temperature $T=0.3$.
\label{fig:nonneut} } 
\end{figure}

We now investigate the robustness of the equivalence between driven and
ageing FD relations with respect to (i)
non-neutrality of the observable $m$, and (ii) changes in driving
mechanism. We start with (i), considering observables for which
$\trapvar =\exp(nE/T)$ (which defines $n$), though still with
$\trapmean =0$.  In this case, the overall amplitude (initial value)
of the correlator depends explicitly on the waiting time (or driving
rate), even in the ageing (or weakly driven) limit. In order to obtain
a limiting FD plot, the correlator and response must be normalised by
the initial value of the correlator. In the driven regime, therefore,
we now plot $\tilde{\chi}(t,\gdot)\equiv \chi(t,\gdot)/C(0,\gdot)$
versus $\tilde{C}(t,\gdot)=C(t,\gdot)/C(0,\gdot)$, with $t$ as the
plotting parameter. In the ageing case, for these non-neutral
observables, care must be taken in constructing the FD plot. For
neutral observables, the usual prescription is to plot $\chi(t,\tw)$
versus $C(t,\tw)$ with $t$ as the plotting parameter.  For non-neutral
observables, however, we see from Eqn.~\ref{eqn:non_eq_fdt} that
the slope of the FD plot is only guaranteed to coincide with $X$ if we
plot $\tilde{\chi}(t,\tw)\equiv \chi(t,\tw)/\chi(t,t)$ versus
$\tilde{C}(t,\tw)\equiv C(t,\tw)/C(t,t)$ {\em with $\tw$ as the
  plotting parameter}. (This coincides with the usual prescription for
neutral observables, as required.)  These normalised FD plots are
shown for $n=0.2$ in Fig.~\ref{fig:nonneut}, and are seen to differ
strongly from each other: equivalence of ageing and driven FD
relations does not hold in the trap model for non-neutral observables.

\begin{figure}[h]
\centerline{\psfig{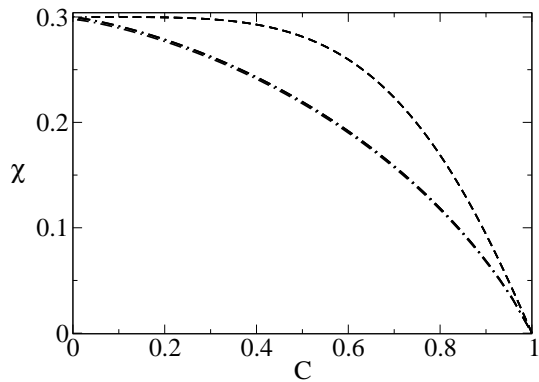}}
\caption{FD plots for the neutral observable with $\trapmean=0$, $\trapvar=1$ for (i) the ageing model with  $t=10^{3},\,10^4,\,10^5,\,10^6$ (dashed lines; different $t$ values indistinguishable) and (ii) a driven model with the alternative driving dynamics of Eqn.~\ref{eqn:drivenB} for $\gdot=10^{-3},\,10^{-4},\,10^{-5},\,10^{-6}$ (dot-dashed lines). The temperature $T=0.3$.
\label{fig:drivenB} } 
\end{figure}

Finally, we consider robustness of the equivalence of ageing and
driven FDT with respect to a change in the driving mechanism. To do
this, we consider a different driving mechanism which adds an
additional hopping process to the trap model whose rate $\gdot$ is
independent of trap depth. This just has the effect of normalizing 
each hopping rate according to $1/\tau\to 1/\tau+\gdot$. The dynamics are now
\be
\partial_t P(E,m,t) = - \left[\frac{1}{\tau(E,m)}+\gdot\right]\,P + Y(t)\,\rho(E)\distrm.
\label{eqn:drivenB}
\ee
with a steady state energy distribution given by
\be
\PEg=\frac{\Yss\rho(E)}{\tau^{-1}(E)+\gdot}.
\ee
The calculation of correlation and response functions is now trivial
since all survival probabilities are simple exponentials; one finds
\bw
\be
\Cg=\int_0^\infty dE \,\trapvar \PEg \exp\left[-\left(\frac{1}{\tau(E)}+\gdot\right)t\right]
\ee
and
\be
\Rg=\int_0^\infty dE \,\trapvar \PEg \frac{1}{1+\gdot\tau(E)}\left\{1-\exp\left[-\left(\frac{1}{\tau(E)}+\gdot\right)t\right]\right\}.
\ee
\ew
The FD plot for the neutral observable $\trapvar=1$ is given in
Fig.~\ref{fig:drivenB}, and seen to differ strongly from the
corresponding FD plot in the undriven model. Hence, the equivalence of
the ageing and driven FD relations is not preserved for this change of
driving mechanism.  However, this is likely to be a consequence of the
fact that this second choice of driving mechanism does not in fact violate
detailed balance, but merely renormalises all the jump rates.

\section{Summary and conclusion}
\label{sec:conc}

In this paper we studied the non-equilibrium FDT in the glass phase of
Bouchaud's trap
model~\cite{Bouchaud92,MonBou96,RinMaaBou00,SolLeqHebCat97}, extended
to incorporate the non-linear driving mechanism of
Ref.~\cite{SolLeqHebCat97}.  After deriving exact expressions for the
correlation and response functions of a generic observable, $m$, we
compared the FD relation for a system driven steadily at rate
$\gdot\to 0$ with that for an ageing system at waiting time $1/\gdot$.
For ``neutral'' observables that are uncorrelated with the system's
average energy, the driven and ageing FD relations are the same, to
within minor logarithmic corrections.  This correspondence does not
apply to non-neutral observables. Finally, we considered an
alternative driving mechanism that renormalises all the hopping rates
according to $1/\tau\to 1/\tau+\gdot$. In this case, the ageing and
driven FD relations differ strongly, even for neutral observables.
Although this is apparently at odds with our central result, it is
likely to result from the fact that this trivial driving mechanism
does not violate detailed balance. Further research is certainly
needed, however, to understand whether other conditions are required 
on driving mechanisms in order to get  steady-state behaviour related to
that of undriven ageing systems.

We return finally to address the fact that the FD relations are
rounded in this model, thus excluding a single effective temperature
within the ageing or driven time sectors (with relaxation times
$O(\tw)$ or $O(\gdot)$ respectively). Similarly `rounded' FDT plots
have recently been found in coarsening models at
criticality~\cite{GodLuc00b}; the limiting value $-X_\infty$ of the
slope for $C\to 0$ was there shown to be a universal amplitude ratio.
It is possible that at least this $X_\infty$ could define a sensible
$\teff$, and in fact both our limiting FDT plots (for the first
driving mechanism) share a common value $X_\infty=0$.  

In conclusion, the FD relation between correlation and response is the
same, to within logarithmic corrections, in the ageing and driven trap
models. We suggest that the limiting value $-X_\infty$ of the slope
for $C\to 0$ could be used to define an effective temperature.
Further work is needed to delineate more fully the class of finite
dimensional driven glassy models that exhibit this behaviour.

Note added: we have recently become aware that Ludovic Berthier has
studied FDT in the driven EA model, and found FD plots the same as in
the ageing model, with agreement between two different observables
(unpublished).

{\bf Acknowledgements}: Financial support from EPSRC (SMF) and the
Nuffield foundation (PS, grant NAL/00361/G) is gratefully
acknowledged. This work was supported in part by the National Science
Foundation under grant number PHY99-07949.

\bibliographystyle{plain}

%\bibliography{ackerson,actin,articles,banding,barham,berrportdecruppe,books,callaghan,chandcits,crystal,crystal_theory,dnatheory,flowcryst,fredrickson,gelbart,helfrich,hsiao,LCtheory,lifshitz,line_tension,malkus,master,mccoy,membs,noirez,notes,onions,phd,phd1,pineweitz,pomeau,poon,pratt,psolutions,rheofolks,ryan,savedrecs.txt,schoot,semenov,sriram,stein,worms2,worms3,worms,zubarev}

\begin{thebibliography}{10}

\bibitem{Barrat98}
A~Barrat.
\newblock {Monte} {Carlo} simulations of the violation of the
  fluctuation-dissipation theorem in domain growth processes.
\newblock {\em Phys.\ Rev.\ E}, 57(3):3629--3632, 1998.

\bibitem{BarBer01}
J~L Barrat and L~Berthier.
\newblock Fluctuation-dissipation relation in a sheared fluid - art. no.
  012503.
\newblock {\em Phys.\ Rev.\ E}, 6301(1):2503--+, 2001.

\bibitem{BerBar02b}
L~Berthier and J~L Barrat.
\newblock Nonequilibrium dynamics and fluctuation-dissipation relation in a
  sheared fluid.
\newblock {\em J.\ Chem.\ Phys.}, 116(14):6228--6242, 2002.

\bibitem{BerBar02}
L~Berthier and J~L Barrat.
\newblock Shearing a glassy material: numerical tests of nonequilibrium
  mode-coupling approaches and experimental proposals.
\newblock {\em Phys.\ Rev.\ Lett.}, 89(9):art. no.--095702, 2002.

\bibitem{BerBarKur00}
L~Berthier, J~L Barrat, and J~Kurchan.
\newblock A two-time-scale, two-temperature scenario for nonlinear rheology.
\newblock {\em Phys.\ Rev.\ E}, 61(5):5464--5472, 2000.

\bibitem{Bouchaud92}
J~P Bouchaud.
\newblock Weak ergodicity breaking and aging in disordered-systems.
\newblock {\em J.\ Phys.\ (France)\ I}, 2(9):1705--1713, 1992.
\newblock The trap model with Gaussian $\rho(E)$, where an equilibrium state exists at all temperatures, had earlier been studied in J~Dyre.
\newblock {\em Phys.\ Rev.\ Lett.}, 58(8): 792--795, 1987.

\bibitem{BouCugKurMez98}
J~P Bouchaud, L~F Cugliandolo, J~Kurchan, and M~M{\'{e}}zard.
\newblock Out of equilibrium dynamics in spin-glasses and other glassy systems.
\newblock In A~P Young, editor, {\em Spin glasses and random fields},
  Singapore, 1998. World Scientific.

\bibitem{BouDea95}
J~P Bouchaud and D~S Dean.
\newblock Aging on {Parisi's} tree.
\newblock {\em J.\ Phys.\ (France)\ I}, 5(3):265--286, 1995.

\bibitem{CorGonLipZan01}
F~Corberi, G~Gonnella, E~Lippiello, and M~Zannetti.
\newblock Effects of an external drive on the fluctuation-dissipation relation
  of phase-ordering systems.
\newblock 2002.
\newblock Preprint cond-mat/0205627.

\bibitem{CriSom87}
A~Crisanti and H~Sompolinsky.
\newblock {\em Phys.\ Rev.\ A}, 36:4922, 1987.

\bibitem{CugKur93}
L~F Cugliandolo and J~Kurchan.
\newblock Analytical solution of the off-equilibrium dynamics of a long- range
  spin-glass model.
\newblock {\em Phys.\ Rev.\ Lett.}, 71(1):173--176, 1993.

\bibitem{CugKur94}
L~F Cugliandolo and J~Kurchan.
\newblock On the out-of-equilibrium relaxation of the
  {Sherrington}-{Kirkpatrick} model.
\newblock {\em J.\ Phys.\ A}, 27(17):5749--5772, 1994.

\bibitem{CugKurLeDouPel97}
L~F Cugliandolo, J~Kurchan, P~Le{D}oussal, and L~Peliti.
\newblock Glassy behaviour in disordered systems with nonrelaxational dynamics.
\newblock {\em Physical Review Letters}, 78(2):350--353, 1997.

\bibitem{CugKurPel97}
L~F Cugliandolo, J~Kurchan, and L~Peliti.
\newblock Energy flow, partial equilibration, and effective temperatures in
  systems with slow dynamics.
\newblock {\em Phys.\ Rev.\ E}, 55(4):3898--3914, 1997.

\bibitem{FieSol02}
S~Fielding and P~Sollich.
\newblock Observable dependence of fluctuation-dissipation relations and
  effective temperatures.
\newblock {\em Phys.\ Rev.\ Lett.}, 8805(5):art. no.--050603, 2002.

\bibitem{GodLuc00b}
C~Godreche and J~M Luck.
\newblock Response of non-equilibrium systems at criticality: ferromagnetic
  models in dimension two and above.
\newblock {\em J.\ Phys.\ A}, 33(50):9141--9164, 2000.
\newblock cond-mat/0001264.

\bibitem{HerGriSol86}
J~A Hertz, Grinstein G, and S~Solla.
\newblock In J~L van Hemmen and I~Morgenstern, editors, {\em Proceedings of the
  Heidelberg Colloquium on Glassy Dynamics and Optimization}, Berlin, 1987.
  Springer Verlag.

\bibitem{Horner96}
H~Horner.
\newblock Drift, creep and pinning of a particle in a correlated random
  potential.
\newblock {\em Zeitschr.\ Phys.\ B}, 100(2):243--257, 1996.

\bibitem{MarParRicRui98}
E~Marinari, G~Parisi, F~Ricci-Tersenghi, and J~J Ruiz-Lorenzo.
\newblock Violation of the fluctuation-dissipation theorem in finite-
  dimensional spin glasses.
\newblock {\em J.\ Phys.\ A-Math.\ Gen.}, 31(11):2611--2620, 1998.

\bibitem{MonBou96}
C~Monthus and J~P Bouchaud.
\newblock Models of traps and glass phenomenology.
\newblock {\em J.\ Phys.\ A}, 29(14):3847--3869, 1996.

\bibitem{Reichl80}
L~E Reichl.
\newblock {\em A modern course in statistical physics}.
\newblock University of Texas Press, Austin, 1980.

\bibitem{RinMaaBou00}
B~Rinn, P~Maass, and J-P Bouchaud.
\newblock Multiple scaling regimes in simple aging models.
\newblock {\em Phys.\ Rev.\ Lett.}, 84(23):5403--5406, 2000.

\bibitem{SolLeqHebCat97}
P~Sollich, F~Lequeux, P~H{\'{e}}braud, and M~E Cates.
\newblock Rheology of soft glassy materials.
\newblock {\em Phys.\ Rev.\ Lett.}, 78:2020--2023, 1997.

\bibitem{Thalmann98}
F~Thalmann.
\newblock {\em Eur.\ Phys.\ J.\ B}, 3:497, 1998.


\end{thebibliography}

\end{document}